\documentclass[a4paper,oneside,final,notitlepage,onecolumn,12pt]{article}%\documentclass[smallextended]{svjour3}
\usepackage{amsmath, amssymb, graphicx}

\begin{document}

%\preprint{}


\author{{Andor Frenkel}\,\thanks{ ~email: frenkel.andor@wigner.mta.hu} %EndAName 
\\ 
\\ 
{Wigner Research Center for Physics, Budapest} }
\title{Dependence of the time-reading process of the Salecker--Wigner quantum clock on the size of the clock}

\maketitle

\begin{abstract}
It is shown in the present note that the degree of the complexity of the time-reading process of the
Salecker--Wigner clock depends on the size of the clock.
This dependence leads to a relation between the size and the accuracy of the clock, and suggests a precise optimal value for the size in agreement with the order of magnitude value established by Salecker and Wigner.
%\keywords{Quantum clock \and Measurement of space--time distances \and Optimal mass}
\end{abstract}

\setcounter{section}{0}
\section{\label{sec:1}Introduction}


According to the laws of classical physics a clock of perfect accuracy is conceivable. 
However, a real clock is a quantum system, the position and the velocity of the center of mass of the hand have indeterminacies, therefore the accuracy of the clock cannot be perfect.
The relation between the accuracy, the running time, the size and the mass of the clock depends on the details of the construction of the clock. 

More than half a century ago H. Salecker and E. P. Wigner proposed a model of a simple quantum clock and of its time-reading device in Section~3 of \cite{1}.
Their clock consists merely of three free quantum bodies one of which is the hand, the two others are the ends of the linear dial.
The reading device comprises a macroscopic time recorder (a classical macroscopic measuring apparatus of von Neumann) and light quanta which scatter with the bodies of the clock and bring the information about the time carried by the clock to the recorder.

The main result in \cite{1} is a formula for the minimal mass $M$ of the bodies of the clock in function of the running time $T$, the accuracy $\tau$ and the linear size $2\ell$ (the distance between the dial bodies) of the clock.
In particular, $M$ turns out to be proportional to the inverse square of $2\ell$.
It is argued in \cite{1} and on page 261 of \cite{2} that it follows from the properties of the clock that the value of $2\ell$ is of the order of $c\tau$.
In the present note it is shown that the complexity of the process of the time-reading depends on the size of the clock and this dependence suggests a precise optimal value of $2\ell$ close to $c\tau$: up to that value, given below in Eq.~\eqref{eq:33} the process of the time-reading is simpler than above it.

A light quantum scattered by a small quantum body spreads in every direction, therefore the probability that a quantum scattered by a body of the clock reaches the recorder is very low.
In order to guarantee that the quanta reach the recorder with certainty the discussion in \cite{1} has been confined to a world with only one space-like dimension.
This restriction is maintained in the present note.
The structure of the Salecker--Wigner (S--W) clock in the real $(1 + 3)$-dimensional world has been investigated in~\cite{3}.

In Sections \ref{sec:2} and \ref{sec:3} below the main features of the (S--W) clock and of its time-reading device are recalled.
In Section~\ref{sec:4} the reasoning suggesting a precise value of the length of the dial is exposed.
The calculations underlying the statements made in Section~\ref{sec:4} are presented in Appendix~2.
Appendix~1 contains the derivation of Eq.~\eqref{eq:4} of the main text.

\section{\label{sec:2}Description of the Salecker--Wigner clock}

\begin{figure}[h]
\includegraphics[width=\hsize]{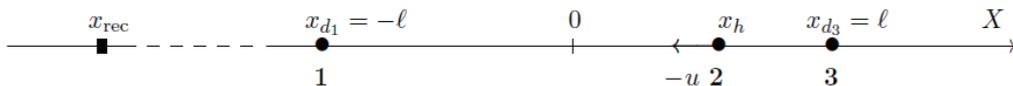}

\caption{\label{fig:1}The coordinates of the three bodies of the clock and of the recorder}
\end{figure}

The S--W clock consists of three free quantum bodies \textbf{1}, \textbf{2}, \textbf{3} of equal masses $M$.
Bodies \textbf{1} and \textbf{3}, at rest in an inertial frame of reference at a distance $2\ell$ from each other are the ends of the linear dial which has no inner material division points.
In conformity with \cite{1} let the centers of mass (c.\ m.) of dial bodies \textbf{1} and \textbf{3} be located at the points $x_{d_1} = -\ell$, $x_{d_3} = \ell$ of the $X$ axis (Fig.~\ref{fig:1}).
(The c.\ m.\ coordinates of the three bodies have quantum indeterminacies.
The exact values refer to the maxima of their Gaussian c.\ m.\ wave packets.)
Body \textbf{2} moving with velocity $-u$ $(0 < u \ll c)$ from body \textbf{3} towards body \textbf{1} is the hand.

It is argued in \cite{1} that the effect of the scatterings of the bodies of the clock with the light quanta of the reading device is negligible as far as the state of motion of these bodies is considered.
The velocity $-u$ of the hand is constant in this approximation and the relation between the length $2\ell$ of the dial and the running time $T$ is
\begin{equation}
\label{eq:1}
2\ell = u T.
\end{equation}
The restriction to non-relativistic velocity $u \ll c$ is in conformity with the use of non-relativistic quantum mechanics in~\cite{1}.

The time $t_c$ carried by the clock is related to the position $x_h$ of the hand.
A simple convenient relation is
\begin{equation}
\label{eq:2}
t_c = \frac{x_h}{-u} = - \frac{x_h}{2\ell} T;
\end{equation}
then the hand starts from dial body $\mathbf 3$ at the initial moment
\begin{equation}
\label{eq:3}
t_c^{in} = - \frac{x_{d_3}}{2\ell} T = - \frac{T}{2},
\end{equation}
passes at the middle $x = 0$ of the dial when $t_c = 0$ and reaches dial body $\mathbf 1$ at the moment $T/2$.

In \cite{1} a slightly different connection between $x_h$ and the time carried by the clock, denoted there by $t_0$ has been adopted.
As shown in Appendix~1 the relation between $t_0$ and $t_c$ is
\begin{equation}
\label{eq:4}
t_0 = (1 + \beta)t_c, \ \ \ \beta = \frac{u}{c} \ll 1.
\end{equation}
Their difference is of no consequence for the results obtained in the present note.

Since the S--W clock is a quantum clock the bodies constituting it have to be treated quantum mechanically.
In conformity with \cite{1} let the c.\ m.\ wave function of the hand be a minimal Gaussian wave packet at the initial moment of time $t_c^{in} = -T/2$.
If the width of the packet, i.e.\ the positional indeterminacy $\Delta x_h$ of the hand at that moment is such that
\begin{equation}
\label{eq:5}
T \lesssim \frac{M(\Delta x_h)^2}{\hbar},
\end{equation}
then during the running time $T$ the width at most doubles, thus $\Delta x_h$ keeps its order of magnitude.
The uncertainty $\tau$ of the time kept by the clock, called the accuracy of the clock in \cite{1}, then is
\begin{equation}
\label{eq:6}
\tau = \frac{\Delta x_h}{u} = \frac{\Delta x_h}{2\ell} T,
\end{equation}
the time interval during which  the c.\ m.\ of the hand is at a given point of space with considerable probability.
\eqref{eq:5} and \eqref{eq:6} lead to the Salecker--Wigner mass formula
\begin{equation}
\label{eq:7}
M \gtrsim \frac{\hbar}{(2\ell)^2} \, \frac{T^3}{\tau^2}.
\end{equation}
(Similarly to \cite{1} many of our formulas are order of magnitude estimates and numerical factors of the order of the unity are often omitted.)

The relative accuracy of the clock is $\tau/T$.
Its inverse is denoted by $n$ in \cite{1};
\begin{equation}
\label{eq:8}
n = \frac{T}{\tau}.
\end{equation}
For a good clock
\begin{equation}
\label{eq:9}
n \gg 1
\end{equation}
should hold.
According to \eqref{eq:6} and \eqref{eq:8} the relation between $\Delta x_h$ and $2\ell$ is
\begin{equation}
\label{eq:10}
\Delta x_h = \frac{2\ell}{n}.
\end{equation}

To estimate the value of $2\ell$ one has to take into account that $2\ell = uT$ and $u \ll c$.
The features of the S--W clock offer only one quantity much smaller than~$1$, namely $1/n$.
Thus the plausible order of magnitude of $u$ is $c/n$, therefore
\begin{equation}
\label{eq:11}
2\ell = uT \approx \frac{c}{n} T = c\tau.
\end{equation}
It is easy to see that the uncertainty of $u$ is still much smaller than~$u$.
Indeed, for a minimal Gaussian wave packet
\begin{equation}
\label{eq:12}
\Delta x_h \cdot \Delta p_h \approx \hbar,
\end{equation}
where
\begin{equation}
\label{eq:13}
p_h = - Mu
\end{equation}
is the momentum of the hand.
It follows from \eqref{eq:12}, \eqref{eq:5}, \eqref{eq:10} and \eqref{eq:1} that
\begin{equation}
\label{eq:14}
\Delta u = \frac{\Delta p_h}{M} \approx \frac{\hbar}{M \Delta x_h} \lesssim \frac{\Delta x_h}{T} \approx \frac{2\ell}{T} \frac{1}{n} = \frac{u}{n} \ll u
\end{equation}
as it should be for a decent clock.

At the time of the publication of \cite{1} and \cite{2} (in the late `50's) atomic clocks of relative accuracy $\tau/T \approx 10^{-13}$  were operating.
Probably this is why in \cite{2} Wigner considers an S--W clock with running time $T \approx 10^{5}$~sec and accuracy $\tau \approx 10^{-8}$~sec, and he notices that such a clock is a macroscopic object: the length $2\ell = c\tau$ of the dial is of the order of a meter, and the mass of the hand, according to \eqref{eq:7}, is $0.1$ gram.
However, as shown in \cite{4}, less accurate S--W clocks can have microscopic size and mass.

\section{\label{sec:3}Description of the time-reading device}

The time-reading device consists of a macroscopic classical time recorder and of light quanta bringing the information about the time carried by the clock to the recorder.
The wave packets of three quanta travel together along the negative $X$ axis toward the clock.
The members of such a triad are denoted by $(\widetilde 1, \widetilde 2, \widetilde 3)$ in this note.
In agreement with \cite{1} each member of a triad is scattered back by a different body of the clock.
To make the bookkeeping easy let dial body $\mathbf 1$ scatter back only the quantum $\widetilde 1$, the hand only the quantum $\widetilde 2$ and dial body $\mathbf 3$ only the quantum $\widetilde 3$.
To achieve this the quanta of a triad must be distinguishable from each other, e.g.\ they should have different colors.

The scattered quanta of a triad, travelling toward the recorder placed at a distant point of the negative $X$ axis are denoted by $(\widehat 1, \widehat 2, \widehat 3)$, with a hat at the place of the tilde of the incoming quanta.

The recorder can register the times of arrival $t_{\widehat 1}^A, t_{\widehat 2}^A, t_{\widehat 3}^A$ of the quanta of a triad.
The time $t_c$ carried by the clock, defined in \eqref{eq:2} above is simply related to the ratio of two differences of these times of arrival.
Namely
\begin{equation}
\label{eq:15}
t_c = \left( - \frac12 + \varrho\right) T
\end{equation}
where
\begin{equation}
\label{eq:16}
\varrho = \frac{t_{\widehat 3}^A - t_{\widehat 2}^A}{t_{\widehat 3}^A - t_{\widehat 1}^A}.
\end{equation}

In order to obtain relation \eqref{eq:15} let us determine the time dependence of the coordinates of the scattered quanta.
The incoming quantum $\widetilde 2$ is scattered back by the hand.
They meet at the point $x_h$ at the moment $t_c$ given in \eqref{eq:2}.
At that moment $\widetilde 2$ becomes $\widehat 2$, and $\widehat 2$ travels toward the recorder with velocity $-c$.
Therefore
\begin{equation}
\label{eq:17}
x_{\widehat 2}(t)  = - c (t - t_c) + x_h, \ \ \ t \geq t_c .
\end{equation}
The incoming quantum $\widetilde 1$ is scattered back by dial body $\mathbf 1$.
They meet at the point $x_{d_1} = -\ell$ earlier than $\widetilde 2$ meets the hand, namely at the moment
\begin{equation}
\label{eq:18}
t_c - \frac{x_h - (-\ell)}{c}.
\end{equation}
Therefore
\begin{align}
x_{\widehat 1}(t) &= -c \left[t - \left(t_c - \frac{x_h + \ell}{c}\right)\right] - \ell \nonumber\\
&= - c(t - t_c) - x_h - 2\ell; \ \ t \geq t_c - \frac{x_h + \ell}{c},
\label{eq:19}
\end{align}
and for $x_{\widehat 3}(t)$ one obtains
\begin{equation}
\label{eq:20}
x_{\widehat 3}(t) = -c (t - t_c) - x_h + 2\ell; \ \ t \geq t_c + \frac{\ell - x_h}{c}.
\end{equation}
Since the three scattered quanta travel with equal velocity $-c$, the differences $x_{\widehat 3} - x_{\widehat 1}$, $x_{\widehat 3} - x_{\widehat 2}$ do not depend on~$t$:
\begin{align}
\label{eq:21}
x_{\widehat 3} - x_{\widehat 1} &= 4\ell,\\
\label{eq:22}
x_{\widehat 3} - x_{\widehat 2} &= 2 (\ell - x_h).
\end{align}
These coordinate differences are simply related to the differences of the times of arrival in \eqref{eq:16}:
\begin{equation}
\label{eq:23}
x_{\widehat 3} - x_{\widehat 1} = c \left(t_{\widehat 3}^A - t_{\widehat 1}^A\right),
\end{equation}
\begin{equation}
\label{eq:24}
x_{\widehat 3} - x_{\widehat 2} = c \left(t_{\widehat 3}^A - t_{\widehat 2}^A\right),
\end{equation}
therefore
\begin{equation}
\label{eq:25}
\varrho = \frac{x_{\widehat 3} - x_{\widehat 2}}{x_{\widehat 3} - x_{\widehat 1}} = \frac12 - \frac{x_h}{2\ell}.
\end{equation}
Remembering that according to \eqref{eq:2}
\begin{equation}
\label{eq:26}
x_h = -2\ell \frac{t_c}{T}
\end{equation}
\eqref{eq:25} gives
\begin{equation}
\label{eq:27}
\varrho = \frac12 + \frac{t_c}{T},
\end{equation}
a relation equivalent to \eqref{eq:15}.

It has been stressed in \cite{1} that it is advantageous to express $t_c$ as a function of $\varrho$ because the ratio of two time intervals, as well as of two parallel distances are Lorentz invariant quantities.
Therefore \eqref{eq:15} is valid even if the dial of the clock is in motion relative to the recorder.
In the present note the discussion will be restricted to the case when both are at rest.
Then
\begin{equation}
\label{eq:28}
t_{\widehat 3}^A - t_{\widehat 1}^A = \frac1c \left(x_{\widehat 3} - x_{\widehat 1}\right) = \frac{4\ell}{c}
\end{equation}
is known, so only $t_{\widehat 2}^A$ and $t_{\widehat 3}^A$ should be registered by the recorder.
$t_c$ is then given as
\begin{equation}
\label{eq:29}
t_c = \left(- \frac12 + \frac{c\left(t_{\widehat 3}^A - t_{\widehat 2}^A\right)}{4\ell} \right) T = \left( - \frac12 + \frac{x_{\widehat 3} - x_{\widehat 2}}{4\ell}\right) T.
\end{equation}

\section{\label{sec:4}Dependence of the process of the time-reading on the length of the dial}

With one triad only a single value of the time carried by the clock can be registered.
In order to make possible more than one reading coming from any of the $n$ time intervals $\tau$ of the running time $T$, in \cite{1} an array of $n$ triads impinges on the clock.
The hand will scatter with a quantum $\widetilde 2$ at regular time intervals $\tau$ if the distance between the consecutive triads travelling toward the clock is
\begin{equation}
\label{eq:30}
(c + u)\tau = c\tau(1 + \beta).
\end{equation}
Here it has been taken into account that the hand is moving with velocity $-u$, opposite to the velocity $c$ of the incoming quanta.
In \cite{1} instead of \eqref{eq:30} the approximate value $c\tau$ has been employed.

In this note the triads travelling toward the clock are denoted by
\begin{equation}
\label{eq:31}
\left(\widetilde 1_k, \widetilde 2_k, \widetilde 3_k\right); \ \ \ k = 1,2,\dots, n,
\end{equation}
and the reflected triads by $\left(\widehat 1_k, \widehat 2_k, \widehat 3_k\right)$.
According to \eqref{eq:29} the scattering of the quantum $\widetilde 2_k$ with the hand occurs at the time
\begin{equation}
\label{eq:32}
t_c = \left( - \frac12 + \frac{c\left(t_{\widehat 3_k}^A - t_{\widehat 2_k}^A \right)}{4\ell} \right) =
\left(- \frac12 + \frac{x_{\widehat 3_k} - x_{\widehat 2_k}}{4\ell}\right) T.
\end{equation}
In this formula the scattered quanta $\widehat 2_k$, $\widehat 3_k$ belong of course to the same triad.
Therefore in addition to the registration of the times of arrival the recorder should recognize the triad partners in the array of the quanta reaching it.
The process of this recognition turns out to depend on the length of the dial.
Below this dependence is outlined.
The underlying calculations are given in Appendix~2.

If the length of the dial is
\begin{equation}
\label{eq:33}
2\ell = \frac12 c\tau(1 + \beta)
\end{equation}
or shorter, then the first quantum $\widehat 3$ arriving at the recorder after a quantum $\widehat 2$ is certainly the triad partner of that $\widehat 2$.
In this case the pairing of the triad partners $(\widehat 2, \widehat 3)$ can be done without the knowledge of the serial number $k$ of the quanta.
However, if the dial is longer the pairing is not so simple.
If the length of the dial is
\begin{equation}
\label{eq:34}
2\ell = c\tau(1 + \beta),
\end{equation}
twice the value given in \eqref{eq:33}, then the triad partner of a $\widehat 2$ is the first $\widehat 3$ arriving after that $\widehat 2$ if
\begin{equation}
\label{eq:35}
k \leq \frac{n + 1}{2}
\end{equation}
and the partner is the second $\widehat 3$ if
\begin{equation}
\label{eq:36}
k > \frac{n + 1}{2}.
\end{equation}
Without information about the value of $k$ the pairing is now ambiguous, causing in the time-reading an ambiguity $T/2$, much larger than the accuracy $\tau$ of the clock.

More generally, if the length of the dial is
\begin{equation}
\label{eq:37}
2\ell = \frac{m}{2} c\tau(1 + \beta)
\end{equation}
with
\begin{equation}
\label{eq:38}
m = 2,3, \dots,
\end{equation}
then depending on the value of $k$ the triad partner of a $\widehat 2$ is one of the $m$ quanta $\widehat 3$ following that~$\widehat 2$.
Without the knowledge of $k$ the ambiguity in the pairing is now $m$-fold and the ambiguities in the time-reading are
\begin{equation}
\label{eq:39}
\frac{1}{m} T, \frac{2}{m} T, \dots, \frac{m - 1}{m} T.
\end{equation}
So, \eqref{eq:33} is the maximal length of the dial at which the time-reading can be carried out without the knowledge of~$k$.
It is therefore convenient to use a clock with such a dial.
\eqref{eq:33} is in agreement with the order of magnitude
\begin{equation}
\label{eq:40}
2\ell \approx c\tau
\end{equation}
proposed in \cite{1} and \cite{2}.

If \eqref{eq:33} holds it is sufficient to switch on the recorder at any moment during the arrival of the quanta and to register the times of arrival of a $\widehat 2$ and of the $\widehat 3$ arriving first after that~$\widehat 2$.
The moment $t_c$ when a quantum $\widetilde 2$ was scattered back by the hand and became the $\widehat 2$ in question is given by the formula
\begin{equation}
\label{eq:41}
t_c = \left( - \frac12 + \frac{c\left(t_{\widehat 3}^A - t_{\widehat 2}^A\right)}{4\ell} \right)T
\end{equation}
in which the value of $k$ is not specified.

If the dial is longer than the value given in \eqref{eq:33}, namely if its length is equal to \eqref{eq:37}, then the knowledge of the serial number $k$ is needed for the identification of the triad partner $\widehat 3$ of a~$\widehat 2$.
Accordingly, the process of the time-reading becomes more complicated, but it is noteworthy that the serial number of a quantum can be obtained without counting the quanta arriving at the recorder.
In particular, as shown in Appendix~2, the serial number of a $\widehat 2$ can be deduced from the difference between its time of arrival at the recorder and the time of arrival of the first quantum~$\widehat 2_1$.
Therefore the recorder should be switched on already before the arrivals of the quanta and the time of arrival of the first quantum $\widehat 2_1$ should be registered.
Then from the difference of the times of arrival of $\widehat 2_1$ and of a $\widehat 2$ registered later the serial number $k$ of that $\widehat 2$ gets known, and its triad partner $\widehat 3_k$ can be identified.
The time $t_c$ when $\widetilde 2_k$ met the hand and became the $\widehat 2_k$ in question is given in \eqref{eq:32}.

%%\appendix
%%\setcounter{section}{0}
\section*{\label{app:1}Appendix 1}

In this appendix relation \eqref{eq:4} between $t_0$ and $t_c$ is derived.

In \cite{1} the time $t_0$ carried by the clock is given in Eq.~\eqref{eq:8}:
\begin{equation}
\label{eq:A1}
t_0 = \frac{c - v}{v} \,\frac{\ell}{c} (2r - 1).
\tag{A1}
\end{equation}
Here $v$ stands for the negative velocity of the hand and
\begin{equation}
\label{eq:A2}
r = \frac{t_{\widehat 2}^A - t_{\widehat 1}^A}{t_{\widehat 3}^A - t_1^A} .
\tag{A2}
\end{equation}
The expression of $v$ and $r$ through the velocity $u$ and the ratio $\varrho$ (see \eqref{eq:16}) used in this note is
\begin{align}
v &= -u,
\label{eq:A3}
\tag{A3}\\
r &= 1 - \varrho.
\label{eq:A4}
\tag{A4}
\end{align}
Substituting \eqref{eq:A3} and \eqref{eq:A4} into \eqref{eq:A1} relation \eqref{eq:4} is easily obtained, with $t_c$ given in \eqref{eq:15}.

\section*{\label{app:2}Appendix 2}

In this appendix it is shown that if the length of the dial is
\begin{equation}
\label{eq:A5}
2\ell = \frac12 c\tau (1 + \beta)
\tag{A5}
\end{equation}
or shorter, then the triad partner of a quantum $\widehat 2$ is the first quantum $\widehat 3$ following that $\widehat 2$ on the way toward the recorder.
In this case the identification of the triad partners $(\widehat 2, \widehat 3)$ can be done without the knowledge of the serial
number~$k$.
However, if the dial is longer than the value in \eqref{eq:A5}, namely if
\begin{equation}
\label{eq:A6}
2\ell = \frac{m}{2} c\tau (1 + \beta), \ \ \ m = 2,3, \dots,
\tag{A6}
\end{equation}
then the pairing depends on the value of~$k$.

The proof of the above statements relies on the comparison of the distance $x_{\widehat 3_k} - x_{\widehat 3_{k - 1}}$ between the consecutive quanta $\widehat 3$ with the distances $x_{\widehat 3_k} - x_{\widehat 2_k}$ between the triad partners $\widehat 2_k$, $\widehat 3_k$.

It is easy to see that
\begin{equation}
\label{eq:A7}
x_{\widehat 3_k} - x_{\widehat 3_{k - 1}} = c\tau (1 + \beta), \ \ \ k = 1,2,\dots, n.
\tag{A7}
\end{equation}
Indeed, as told before Eq.\ \eqref{eq:30}, $c\tau(1 + \beta)$ is the distance between the consecutive triads impinging on the clock, in particular between the consecutive quanta $\widetilde 3$ scattered back by dial body $\mathbf 3$ at rest.
The scattered $\widetilde 3$'s are the $\widehat 3$'s, therefore the distance between the consecutive $\widehat 3$'s is equal to the same distance $c\tau (1 + \beta)$.

Let us now look at the distances $x_{\widehat 3_k} - x_{\widehat 2_k}$.
According to \eqref{eq:22}
\begin{equation}
\label{eq:A8}
x_{\widehat 3_k} - x_{\widehat 2_k} = 2\left(\ell - x_h^{(k)}\right), \ \ \ k = 1,2,\dots, n,
\tag{A8}
\end{equation}
where $x_h^{(k)}$ is the coordinate of the hand at the moment of its scattering with the quantum $\widetilde 2_k$ which becomes $\widehat 2_k$ at that very moment.
As told before Eq.~\eqref{eq:30} such scatterings follow each other at time intervals~$\tau$.
During one interval the hand covers the distance
\begin{equation}
\label{eq:A9}
u\tau = \frac{2\ell}{T} \tau = \frac{2\ell}{n}
\tag{A9}
\end{equation}
in the negative $X$ direction, therefore the relation between $x_h^{(k)}$ and $x_h^{(1)}$ is
\begin{equation}
\label{eq:A10}
x_h^{(k)} = x_h^{(1)} - (k - 1) \frac{2\ell}{n}, \ \ \ k = 1,2,\dots, n.
\tag{A10}
\end{equation}
Since the hand starts from the point $x_{d_3} = \ell$, its scattering with the first quantum $\widetilde 2_1$ occurs in the interval 
$\left[ \ell, \ell - \frac{2\ell}{n}\right]$.
Let us consider the case when this scattering happens at the middle point
\begin{equation}
\label{eq:A11}
x_h^{(1)} = \ell - \frac{\ell}{n}
\tag{A11}
\end{equation}
of the interval.
The results obtained in this special case concerning the time-reading process are valid also for other points of the interval.
The mathematical derivation involves cumbersome notation in the general case.

From \eqref{eq:A8}, \eqref{eq:A10} and \eqref{eq:A11} it follows that
\begin{equation}
\label{eq:A12}
x_{\widehat 3_k} - x_{\widehat 2_k} = 4\ell \frac{k - \frac12}{n}, \ \ \ k = 1,2, \dots, n.
\tag{A12}
\end{equation}

\begin{figure}[h]
\includegraphics[width=\hsize]{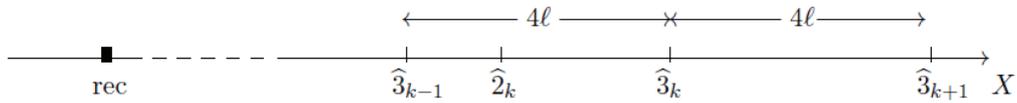}

\caption{\label{fig:2}The case $2\ell = \frac12 c\tau(1 + \beta)$}
\end{figure}

Let us now look at the case when the length of the dial is given by \eqref{eq:A5}:
\begin{equation}
\label{eq:A13}
2\ell = \frac12 c\tau (1 + \beta).
\tag{A13}
\end{equation}
Then according to \eqref{eq:A7}
\begin{equation}
\label{eq:A14}
x_{\widehat 3_k} - x_{\widehat 3_{k - 1}} = 4\ell,
\tag{A14}
\end{equation}
and \eqref{eq:A12} leads to
\begin{equation}
\label{eq:A15}
x_{\widehat 3_k} - x_{\widehat 2_k} = \left(x_{\widehat 3_k} - x_{\widehat 3_{k - 1}}\right) \frac{k - \frac12}{n} < x_{\widehat 3_k} - x_{\widehat 3_{k - 1}}
\tag{A15}
\end{equation}
for all values of~$k$.
\eqref{eq:A15} says that $\widehat 2_k$ travels toward the recorder between $\widehat 3_{k - 1}$ and $\widehat 3_k$ (Fig.~\ref{fig:2}), so the triad partner of $\widehat 2_k$ is indeed the first quantum $\widehat 3$ following $\widehat 2_k$.

Let us now consider the cases when the length of the dial is given by relation \eqref{eq:A6}.
The general result for $m \geq 2$ described in Section~\ref{sec:4} can be inferred from the special cases $m = 2$ and $m = 3$ discussed below.

\begin{figure}[h]
\includegraphics[width=\hsize]{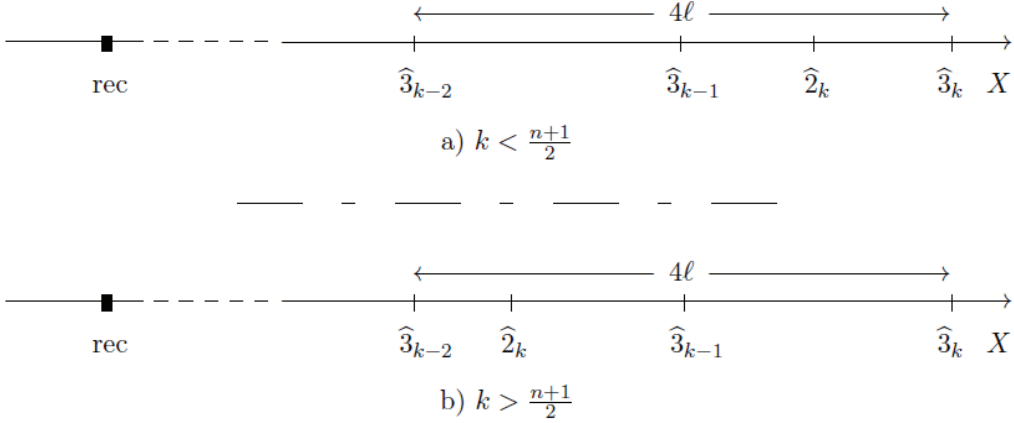}

\caption{\label{fig:3}The cases $m = 2$}
\end{figure}

For $m = 2$ \eqref{eq:A6} and \eqref{eq:A7} give
\begin{equation}
\label{eq:A16}
x_{\widehat 3_k} - x_{\widehat 3_{k - 1}} = 2\ell,
\tag{A16}
\end{equation}
so \eqref{eq:A12} becomes
\begin{equation}
\label{eq:A17}
x_{\widehat 3_k} - x_{\widehat 2_k} = \left(x_{\widehat 3_k} - x_{\widehat 3_{k - 1}}\right) \frac{2k - 1}{n}.
\tag{A17}
\end{equation}
According to this relation if
\begin{equation}
\label{eq:A18}
k < \frac{n + 1}{2}
\tag{A18}
\end{equation}
then
\begin{equation}
\label{eq:A19}
x_{\widehat 3_k} - x_{\widehat 2_k} < x_{\widehat 3_k} - x_{\widehat 3_{k - 1}},
\tag{A19}
\end{equation}
$\widehat 2_k$ travels between $\widehat 3_{k - 1}$ and $\widehat 3_k$, thus the triad partner of $\widehat 2_k$ is again the first $\widehat 3$ following $\widehat 2_k$ (Fig.~\ref{fig:3}a).

If
\begin{equation}
\label{eq:A20}
k = \frac{n + 1}{2}
\tag{A20}
\end{equation}
then $\widehat 2_k$ and $\widehat 3_{k - 1}$ travel together, and the triad partner of $\widehat 2_k$ is still the first $\widehat 3$ following $\widehat 2_k$.
However, if
\begin{equation}
\label{eq:A21}
k > \frac{n + 1}{2}
\tag{A21}
\end{equation}
then
\begin{equation}
\label{eq:A22}
x_{\widehat 3_k} - x_{\widehat 2_k} > x_{\widehat 3_k} - x_{\widehat 3_{k - 1}},
\tag{A22}
\end{equation}
$\widehat 2_k$ travels between $\widehat 3_{k - 2}$ and $\widehat 3_{k - 1}$, and the triad partner of $\widehat 2_k$ is the second $\widehat 3$ following $\widehat 2_k$.

Dealing with relations involving $k$ and $n$ one should take into account that both $k$ and $n$ are integers.
Thus for even values of $n$ the case \eqref{eq:A20} does not occur, and in \eqref{eq:A18} the highest occurring value of $k$ is $n/2$.

If the recorder is turned on at a moment during the arrival of the quanta, the serial number $k$ of a registered quantum $\widehat 2$ is not known, and it is ambiguous whether the first or the second quantum $\widehat 3$ is its triad partner.
According to \eqref{eq:A16} the distance between these $\widehat 3$'s is $2\ell$, and from \eqref{eq:32} it follows that the ambiguity in $t_c$ is
\begin{equation}
\label{eq:A23}
\Delta t_c = \frac{2\ell}{4\ell} T = \frac{T}{2},
\tag{A23}
\end{equation}
much larger than the accuracy $\tau$ of the clock.

Let us now look at the case $m = 3$.
Then according to \eqref{eq:A6} and \eqref{eq:A7}
\begin{equation}
\label{eq:A24}
x_{\widehat 3_k} - x_{\widehat 3_{k - 1}} = \frac{4\ell}{3}, \ \ \ k = 1,2,\dots, n,
\tag{A24}
\end{equation}
and with \eqref{eq:A12}
\begin{equation}
\label{eq:A25}
x_{\widehat 3_k} - x_{\widehat 2_k} = \left(x_{\widehat 3_k} - x_{\widehat 3_{k - 1}}\right) 3 \frac{k - \frac12}{n}.
\tag{A25}
\end{equation}
The triad partner of $\widehat 2_k$ is now the first $\widehat 3$ following $\widehat 2_k$ if
\begin{equation}
\label{eq:A26}
x_{\widehat 3_k} - x_{\widehat 2_k} \leq x_{\widehat 3_k} - x_{\widehat 3_{k - 1}},
\tag{A26}
\end{equation}
i.e.\ according to \eqref{eq:A25} if
\begin{equation}
\label{eq:A27}
k \leq \frac{n}{3} + \frac12.
\tag{A27}
\end{equation}
The triad partner of $\widehat 2_k$ is the second $\widehat 3$ following $\widehat 2_k$ if
\begin{equation}
\label{eq:A28}
x_{\widehat 3_k} - x_{\widehat 3_{k - 1}} < x_{\widehat 3_k} - x_{\widehat 2_k} \leq x_{\widehat 3_k} - x_{\widehat 3_{k - 2}} = 2 \left(x_{\widehat 3_k} - x_{\widehat 3_{k - 1}}\right),
\tag{A28}
\end{equation}
i.e.\ if
\begin{equation}
\label{eq:A29}
\frac{n}{3} + \frac12 < k \leq \frac{2n}{3} + \frac12,
\tag{A29}
\end{equation}
and the triad partner is the third $\widehat 3$ if
\begin{equation}
\label{eq:A30}
\frac{2n}{3} + \frac12 < k \leq n + \frac12.
\tag{A30}
\end{equation}

If the value of $k$ is not registered by the recorder the ambiguity in the choice of the triad partner $\widehat 3$ of a $\widehat 2$ is now three-fold.
According to \eqref{eq:A24} the distance between $\widehat 3_k$ and $\widehat 3_{k - 1}$, as well as between $\widehat 3_{k - 1}$ and $\widehat 3_{k - 2}$ is $4\ell / 3$, between $\widehat 3_k$ and $\widehat 3_{k - 2}$ it is $8\ell / 3$.
Therefore from \eqref{eq:32} it follows that the  ambiguities in the value of $t_c$ are $T/3$ and $2T/3$.

% If in two-column mode, this environment will change to single-column
% format so that long equations can be displayed. Use
%\begin{widetext}
%\end{widetext}

%\begin{acknowledgments}
%\end{acknowledgments}

\bibliography{basename of .bib file}

\end{document}